\documentclass[twocolumn,showpacs,amsmath,amssymb,10pt]{revtex4}
\usepackage{dcolumn}

\usepackage[dvips]{graphicx}
\usepackage{longtable}
\usepackage{stmaryrd}

\begin{document}

\title{A generalized exchange-correlation functional: the Neural-Networks approach}
\author{Xiao Zheng, LiHong Hu, XiuJun Wang, and GuanHua Chen}
\affiliation{Department of Chemistry, The University of Hong Kong,
Hong Kong, China}

\date{\today}

\begin{abstract}
A Neural-Networks-based approach is proposed to construct a new type of
exchange-correlation functional for density functional theory. It is applied to improve
B3LYP functional by
taking into account of high-order contributions to the exchange-correlation functional.
The improved B3LYP functional is based on a neural network whose structure and synaptic weights are
determined from 116 known experimental atomization energies, ionization potentials,
proton affinities or total atomic energies which were used by Becke in his pioneer work on
the hybrid functionals [J. Chem. Phys. ${\bf 98}$, 5648~(1993)].
It leads to better agreement between the
first-principles calculation results and these 116 experimental data.
The new B3LYP functional is further tested by applying it to
calculate the ionization potentials of 24 molecules of the G2 test set.
The 6-311+G(3{\it df},2{\it p}) basis set is employed in the calculation,
and the resulting root-mean-square error is reduced to 2.2 kcal$\cdot$mol$^{-1}$ in
comparison to 3.6 kcal$\cdot$mol$^{-1}$ of conventional B3LYP/6-311+G(3{\it df},2{\it p})
calculation.
\end{abstract}

\maketitle

\section{Introduction}
\label{intro}

Density functional theory (DFT) converts many-electron problems into effective
one-electron problems. This conversion is rigorous if the exact exchange-correlation
functional is known. It is thus important to find the accurate DFT
exchange-correlation functionals. Much progress has
been made, primarily due to the development of generalized gradient
approximation (GGA)~\cite{b88,lyp,pw91} and hybrid functionals~\cite{b3lyp}.
Existing exchange-correlation functionals include local or nearly local
contributions such as local spin density approximation (LSDA)~\cite{ks}
and GGA~\cite{b88,lyp,pw91}, and nonlocal terms, for instance, exact exchange
functional. Although these local and nonlocal terms
account for the bulk contributions to exact exchange-correlation functional,
high-order contributions are yet to be identified and taken into account.
Conceding that it is exceedingly difficult to derive analytically the exact
universal exchange-correlation functional, we resort to an entirely different
approach.

An important methodology in the development of exchange-correlation
functionals has been established by utilizing highly accurate
experimental data to determine exchange-correlation functionals~\cite{b3lyp,b97,handy}.
Becke pioneered this semiempirical approach and determined
the three parameters in B3LYP functional~\cite{g98} by
a least-square fit to 116 molecular and atomic energy data~\cite{b3lyp}.
Building upon this semiempirical methodology, we propose here a new
approach which takes into account of high-order contributions beyond the
existing local and nonlocal exchange-correlation functionals.

Since its beginning in the late fifties, Neural Networks has been
applied to various engineering problems, such as robotics, pattern
recognition, and speech~\cite{PRNN}. A neural network is a highly
nonlinear system, and is suitable to determine or mimic the
complex relationships among relevant physical variables. Recently
we developed a combined first principles calculation and
Neural-Networks correction approach to improve significantly the
accuracy of calculated thermodynamic properties~\cite{lhhu}. In
this work, we develop a Neural-Networks-based approach to
construct the DFT exchange-correlation functional and apply it to
improve the results of the popular B3LYP calculations. In Section
II we describe the Neural-Networks-based methodology and report
our work leading to improved B3LYP calculations. The results of
the improved B3LYP calculations and their comparisons to the
experimental data are given in Section III. Further discussion is
given in Section IV.

\section{Methodology}
\label{method}

B3LYP functional
is a hybrid functional composed of several local and nonlocal exchange and correlation
contributions, and can be expressed as
\begin{widetext}
\begin{eqnarray}
E_{XC} = a_0 E_X^{Slater} + (1-a_0) E_X^{HF} + a_X \Delta E_X^{Becke}
       + a_C E_C^{LYP} + (1-a_C) E_C^{VMN},\label{eq:one}
\end{eqnarray}
\end{widetext}
where E$_X^{Slater}$ is the local spin density exchange functional~\cite{hk,ks,lsda}, E$_X^{HF}$
is the exact exchange functional, E$_X^{Becke}$ is Becke's gradient-corrected exchange
functional~\cite{b88}, E$_C^{LYP}$ is the correlation functional of Lee, Yang and Parr~\cite{lyp},
and E$_C^{VMN}$ represents the correlation functional proposed by Vosko, Wilk and Nusair~\cite{vmn}.
The values of its three parameters, $a_0$, $a_X$ and $a_C$,
dictate the contributions of various terms. They
have been determined via the least-square fit to the 116 atomization
energies (AEs), ionization potentials (IPs), proton affinities (PAs) and total atomic
energies (TAEs) by Becke~\cite{b3lyp}, and are 0.80, 0.72 and 0.81, respectively.
Note that $a_X$$<$$a_0$$<$$a_C$.
B3LYP functional explicitly consists of the first and second rungs of the Jacob's ladder of
density functional approximation~\cite{perdew00} and the partial exact exchange functional~\cite{b3lyp}.
Being determined via the least-square fit to the 116 experimental data, B3LYP functional includes
implicitly the high-order contributions to the exact functional such as those in the meta-GGA
functional~\cite{perdew00}. These high-order contributions are averaged over the 116 energy data
~\cite{b3lyp}, and their functional forms or the values of $a_0$, $a_X$ and $a_C$
are assumed invariant for all types of atomic or molecular systems. Since high-order
contributions to the exact exchange-correlation energy are in fact system-dependent,
their inclusion in Eq.~(\ref{eq:one})
leads to the system-dependence of a$_0$, a$_X$ and a$_C$ which is in turn dictated by the
characteristic properties of the system. The challenge is to identify these characteristic properties,
and more importantly, to determine their quantitative relationships to the values of a$_0$, a$_X$ and
a$_C$. These characteristic properties, termed as the physical descriptors of the system, satisfy two
criteria: (1) they must be of purely electronic nature, since the
exact exchange-correlation functional is a universal functional of electron density
only; and (2) they should reflect the electron distribution.
After identifying these physical descriptors that are related to the high-order contributions
to the exchange-correlation functional, we employ Neural Networks to determine their quantitative
relationships to a$_0$, a$_X$ and a$_C$. Instead of being taken as a system-dependent semiempirical
functional, the resulting neural network can be viewed as a generalized universal exchange-correlation
functional. It can be systematically improved upon the availability of new experimental data.

Beyond the GGA, Perdew and co-workers~\cite{p99} proposed the meta-GGA in which
the exchange-correlation functional depends explicitly on the kinetic energy
density of the occupied Kohn-Sham orbitals,
\begin{eqnarray}
\tau({\bf r}) = {1\over2} \sum_\alpha^{occ} |\nabla\psi_\alpha({\bf r})|^2
\end{eqnarray}
where $\psi_\alpha({\bf r})$ is the wave function of an occupied Kohn-Sham
orbital $\alpha$. The total kinetic energy of the electronic system,
${\cal T} = \int\tau({\bf r})d^3{\bf r}$, should relate closely to the
high-order contributions to B3LYP functional, and is thus chosen as a key physical descriptor.
The exchange-correlation functional is uniquely determined by the electron
density distribution $\rho({\bf r})$.
$\rho({\bf r})$ can be expanded in terms of the multipole moments.
Being the zeroth-order term of the expansion, the total number of
electrons $N_t$ is recognized as a natural physical descriptor, and
the dipole and quadrupole moments of the system are selected as another two descriptors.
We use the magnitude of the dipole moment $D\equiv\sqrt{d_x^2+d_y^2+d_z^2}$ for the
dipole descriptor, where $d_i~(i=x,y,z)$ is a component of the dipole vector.
For the quadrupole descriptor, we choose $Q\equiv\sqrt{Q_{xx}^2+Q_{yy}^2+Q_{zz}^2}$, where $Q_{ii}~
(i=x,y,z)$ is a diagonal element of the quadrupole tensor. The exchange functional accounts for the
exchange interaction among the electrons of the same spin. Spin multiplicity $g_S$ is
thus adopted as a physical descriptor as well.

Our neural network adopts a three-layer architecture which consists of an
input layer, a hidden layer and an output layer~\cite{PRNN}. The values of the physical
descriptors, $g_S$, $N_t$, D, ${\cal T}$ and Q, are inputted into the
neural network at the input layer. The values of the modified $a_0$, $a_X$ and
$a_C$ for each atom or molecule, denoted by $\tilde{a}_0$, $\tilde{a}_X$ and $\tilde{a}_C$,
are obtained at the output layer.
Different layers are connected via the synaptic weights~\cite{PRNN}.
The neural network structure such as the number of hidden neurons
at the hidden layer is to be determined.

We take the 116 experimental energies that were employed by Becke~\cite{b3lyp} as our
training set, and they are utilized to determine the structure of our neural network
and its synaptic weights.
Instead of the basis-set-free calculations carried out by Becke~\cite{b3lyp},
we adopt a Gaussian-type-function (GTF) basis set, 6-311+G(3{\it df},2{\it p}),
in our calculations. Geometry of every molecule is optimized directly using
conventional B3LYP/6-311+G(3{\it df},2{\it p}). The values of $\cal T$,
D and Q are obtained at the same level of calculations.
Besides $g_S$, $N_t$, D, ${\cal T}$ and Q, a bias is introduced as another input and its
value is set to $1$ in all cases. The values of $\tilde{a}_0$, $\tilde{a}_X$ and $\tilde{a}_C$
vary from system to system, and are used to modify the B3LYP functional
for each atom or molecule. The modified B3LYP functional is subsequently used to
evaluate its AE, IP, PA, or TAE. The resulting energies are then compared
to their experimental counterparts, and the comparison is used to tune the synaptic
weights of our neural network. The process is iterated until the differences between
the calculated and measured energies are small enough for all the molecules or atoms in
the training set, and the neural network is then considered as converged, {\it i.e.},
its synaptic weights are determined.
 \begin{figure*}
  \includegraphics[scale=0.80]{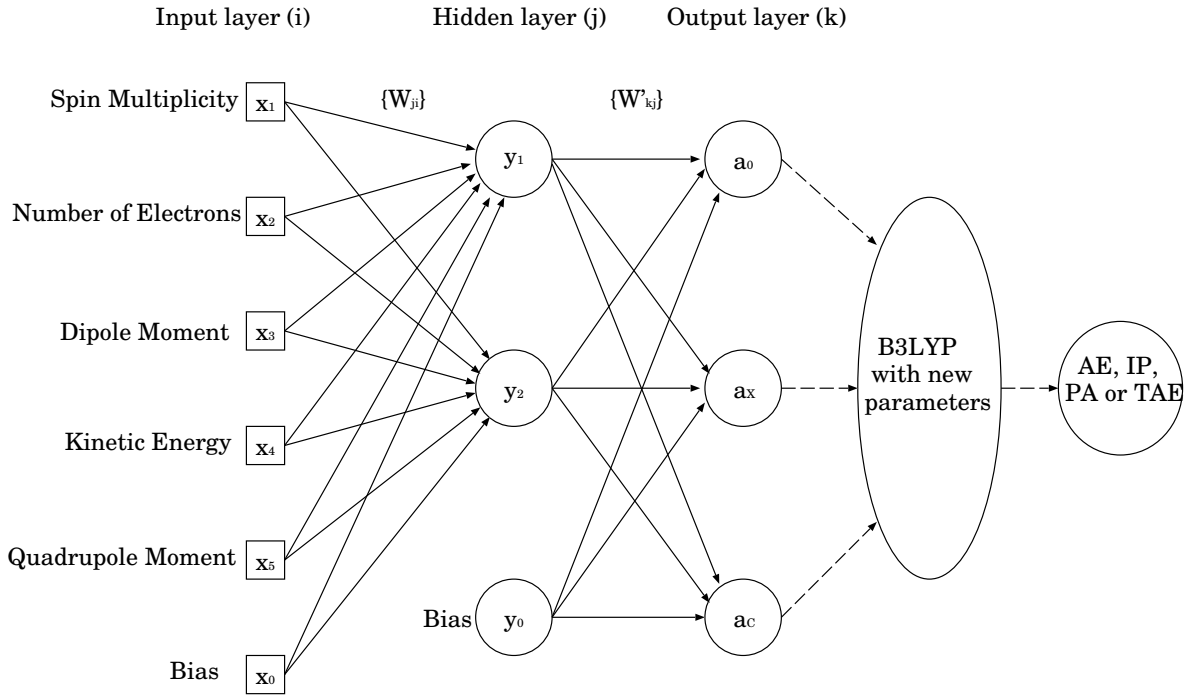}
  \label{fig.1}
  \caption{Architectural graph of our neural network and flow chart of our calculations}
  \end{figure*}

The structure and synaptic weights of our neural network are
optimized via a cross-validation technique~\cite{rbfnn}. The 116
energy values are randomly partitioned into six subsets of equal
size. Five of them are used to train the weights of the neural
network, and are termed as the estimation subset. The sixth is
used to compare the prediction of current neural network, and is
termed as the validation subset. This procedure is repeated six
times in rotation to assess the performance of current neural
network. The number of neurons in the hidden layer is varied from
1 to 5 to decide the optimal structure of the neural network. We
find that the hidden layer containing two neurons yields the best
overall results, {\it i.e.}, the minimal root-mean-square (RMS)
errors and the minimal RMS difference between the estimation and
validation subsets (less than 0.2 kcal$\cdot$mol$^{-1}$).
Minimizing the RMS difference between the estimation and
validation subsets helps ensure the predictive capability of our
neural network. Therefore, the 6-3-3 structure is adopted for our
neural network, see Fig. ~\ref{fig.1}. The input values at the
input layer, $x_1$, $x_2$, $x_3$, $x_4$, $x_5$ and $x_0$ are
$g_S$, $N_t$, D, ${\cal T}$, Q and bias, respectively. Except for
the bias, input values are scaled before being inputted into the
neural network as follows,
\begin{eqnarray}
x_i &=& (C_1-C_2){\cdot}p_i+C_2{\cdot}p^{max}_i-C_1{\cdot}p^{min}_i \over p^{max}_i - p^{min}_i\
\end{eqnarray}
where $C_1$ and $C_2$ are two constants between 0 and 1 that set the upper and lower boundaries,
$p_i$ and $x_i$ are the values of the physical descriptor before and after the
scaling, and $p^{max}_i$ and $p^{min}_i$ are the maximum and
minimum values of the descriptor ($i$=1-5). In our neural network we adopt $C_1 = 0.9$ and $C_2 = 0.1$,
therefore all the inputs {$x_i$} are within the interval [0.1, 0.9].
The biases are introduced at both the input and hidden layers and their value are set to unity.
The synaptic weights $\{ W_{ji}\}$ connect the input layer $\{x_i\}$ and the hidden
neurons $\{y_j\}$, and $\{ W'_{kj} \}$ connect the hidden neurons and the output.
The corrected $\tilde{a}_0$, $\tilde{a}_X$ and $\tilde{a}_C$ are given at the output layer,
and they are related to the input $\{x_i\}$ as
\begin{eqnarray}
\tilde{a}_0 &=& Sigb\{[\sum_{j=1}^{2}W'_{1j}{\cdot}Siga ( \sum_{i=0}^{5}W_{ji} x_i )]+W'_{10}   \}\\
\tilde{a}_X &=& Sigb\{[\sum_{j=1}^{2}W'_{2j}{\cdot}Siga ( \sum_{i=0}^{5}W_{ji} x_i )]+W'_{20}   \}\\
\tilde{a}_C &=& Sigb\{[\sum_{j=1}^{2}W'_{3j}{\cdot}Siga ( \sum_{i=0}^{5}W_{ji} x_i )]+W'_{30}   \}
\end{eqnarray}
where $Siga(v) = {1\over 1+exp(-\alpha v)}$ and $Sigb(v) = {\beta tanh(\gamma v)}$, and
$\alpha$ and $\gamma$ are the parameters that control the switch steepness
of Sigmoidal functions $Siga(v)$ and $Sigb(v)$. An error back-propagation learning
procedure~\cite{nature533} is used to optimize the values of
$W_{ji}$ and $W'_{kj}$ ($i$=0-5, and $j$=0-2. Zero indices are referred
to the biases).

\section{Results}

\begin{longtable*}{@{\extracolsep{0.1in}}p{0.2in}p{0.4in}p{0.4in}p{0.4in}p{0.4in}p{0.4in}p{0.4in}p{0.4in}p{0.4in}p{0.4in}}
\caption{Descriptors and Parameters of Training Set}\\
 \hline \hline
 \\
 No. & Name & g$_S$ & Nt &~D~ & $\cal T$~~ &  Q~~~ & $\tilde{a}_0$ & $\tilde{a}_X$ & $\tilde{a}_C$ \\
            &      &     &    &~(DB)   & ~~(a.u.) &  ~~(DB$\cdot$\AA)  &       &       &       \\
 \hline \endfirsthead
 \multicolumn{10}{c}{Descriptors and Parameters of Training Set continued...}\\
\hline
& \\
No. & Name & g$_S$ & Nt &~D~ & $\cal T$~~ &  Q~~~ & $\tilde{a}_0$ & $\tilde{a}_X$ & $\tilde{a}_C$ \\
            &      &     &    &~(DB)   & ~~(a.u.) &  ~~(DB$\cdot$\AA)  &       &       &       \\
\hline
& \\
 \endhead 

  1 & H$_2$                      & 1 &  2 & 0.00 &   1.83 &  3.35 & 0.779 & 0.726 & 0.906 \\
  2 & LiH                        & 1 &  4 & 5.72 &   8.98 & 10.59 & 0.788 & 0.737 & 0.911 \\
  3 & BeH                        & 2 &  5 & 0.29 &  16.73 & 14.41 & 0.767 & 0.722 & 0.927 \\
  4 & CH                         & 2 &  7 & 1.48 &  41.17 & 12.51 & 0.771 & 0.726 & 0.927 \\
  5 & CH$_2$($^3$B$_1$)          & 3 &  8 & 0.61 &  45.29 & 12.98 & 0.752 & 0.714 & 0.939 \\
  6 & CH$_2$($^1$A$_1$)          & 1 &  8 & 1.81 &  44.98 & 13.80 & 0.789 & 0.737 & 0.909 \\
  7 & CH$_3$                     & 2 &  9 & 0.00 &  49.28 & 13.80 & 0.771 & 0.727 & 0.927 \\
  8 & CH$_4$                     & 1 & 10 & 0.00 &  53.68 & 14.70 & 0.789 & 0.737 & 0.908 \\
  9 & NH                         & 3 &  8 & 1.54 &  58.63 & 10.93 & 0.753 & 0.715 & 0.939 \\
 10 & NH$_2$                     & 2 &  9 & 1.82 &  63.23 & 12.10 & 0.773 & 0.729 & 0.927 \\
 11 & NH$_3$                     & 1 & 10 & 1.53 &  68.24 & 13.14 & 0.791 & 0.739 & 0.909 \\
 12 & OH                         & 2 &  9 & 1.68 &  79.89 & 10.08 & 0.774 & 0.729 & 0.927 \\
 13 & OH$_2$                     & 1 & 10 & 1.91 &  85.35 & 11.07 & 0.791 & 0.740 & 0.909 \\
 14 & FH                         & 1 & 10 & 1.85 & 105.36 &  9.13 & 0.792 & 0.740 & 0.908 \\
 15 & Li$_2$                     & 1 &  6 & 0.00 &  16.62 & 22.29 & 0.786 & 0.735 & 0.911 \\
 16 & LiF                        & 1 & 12 & 6.22 & 116.17 & 10.65 & 0.797 & 0.746 & 0.911 \\
 17 & C$_2$H$_2$                 & 1 & 14 & 0.00 & 101.73 & 20.75 & 0.794 & 0.743 & 0.910 \\
 18 & C$_2$H$_4$                 & 1 & 16 & 0.00 & 111.54 & 23.65 & 0.796 & 0.746 & 0.910 \\
 19 & C$_2$H$_6$                 & 1 & 18 & 0.00 & 121.45 & 26.44 & 0.798 & 0.748 & 0.911 \\
 20 & CN                         & 2 & 13 & 1.38 & 111.45 & 18.85 & 0.779 & 0.736 & 0.928 \\
 21 & HCN                        & 1 & 14 & 3.04 & 117.02 & 19.43 & 0.797 & 0.747 & 0.911 \\
 22 & CO                         & 1 & 14 & 0.10 & 135.49 & 19.13 & 0.795 & 0.744 & 0.910 \\
 23 & HCO                        & 2 & 15 & 1.67 & 140.01 & 19.86 & 0.782 & 0.739 & 0.928 \\
 24 & H$_2$CO                    & 1 & 16 & 2.41 & 145.41 & 20.69 & 0.799 & 0.748 & 0.911 \\
 25 & H$_3$COH                   & 1 & 18 & 1.69 & 155.46 & 22.83 & 0.800 & 0.750 & 0.911 \\
 26 & N$_2$                      & 1 & 14 & 0.00 & 132.87 & 18.79 & 0.795 & 0.744 & 0.909 \\
 27 & H$_2$NNH$_2$               & 1 & 18 & 1.93 & 152.86 & 22.79 & 0.800 & 0.750 & 0.911 \\
 28 & NO                         & 2 & 15 & 0.14 & 155.35 & 18.57 & 0.781 & 0.737 & 0.927 \\
 29 & O$_2$                      & 3 & 16 & 0.00 & 178.05 & 17.78 & 0.766 & 0.727 & 0.939 \\
 30 & HOOH                       & 1 & 18 & 0.00 & 187.79 & 19.53 & 0.799 & 0.748 & 0.909 \\
 31 & F$_2$                      & 1 & 18 & 0.00 & 229.74 & 16.40 & 0.800 & 0.749 & 0.909 \\
 32 & CO$_2$                     & 1 & 22 & 0.00 & 246.26 & 28.63 & 0.804 & 0.754 & 0.912 \\
 33 & SiH$_2$($^1$A$_1$)         & 1 & 16 & 0.09 & 300.05 & 27.42 & 0.802 & 0.753 & 0.913 \\
 34 & SiH$_2$($^3$B$_1$)         & 3 & 16 & 0.07 & 300.26 & 27.00 & 0.773 & 0.735 & 0.940 \\
 35 & SiH$_3$                    & 2 & 17 & 0.00 & 306.45 & 28.06 & 0.790 & 0.747 & 0.930 \\
 36 & SiH$_4$                    & 1 & 18 & 0.00 & 312.63 & 28.84 & 0.803 & 0.754 & 0.913 \\
 37 & PH$_2$                     & 2 & 17 & 0.52 & 353.40 & 26.07 & 0.791 & 0.749 & 0.930 \\
 38 & PH$_3$                     & 1 & 18 & 0.57 & 360.14 & 27.16 & 0.805 & 0.756 & 0.913 \\
 39 & SH$_2$                     & 1 & 18 & 1.00 & 411.71 & 25.00 & 0.806 & 0.757 & 0.914 \\
 40 & ClH                        & 1 & 18 & 1.12 & 466.75 & 22.63 & 0.807 & 0.758 & 0.914 \\
 41 & Na$_2$                     & 1 & 22 & 0.00 & 344.67 & 33.68 & 0.807 & 0.758 & 0.914 \\
 42 & Si$_2$                     & 3 & 28 & 0.00 & 625.88 & 46.53 & 0.796 & 0.760 & 0.942 \\
 43 & P$_2$                      & 1 & 30 & 0.00 & 744.63 & 45.05 & 0.817 & 0.771 & 0.919 \\
 44 & S$_2$                      & 3 & 32 & 0.00 & 866.42 & 44.07 & 0.804 & 0.768 & 0.942 \\
 45 & Cl$_2$                     & 1 & 34 & 0.00 & 994.10 & 42.53 & 0.821 & 0.775 & 0.920 \\
 46 & NaCl                       & 1 & 28 & 8.74 & 662.67 & 30.03 & 0.817 & 0.772 & 0.919 \\
 47 & SiO                        & 1 & 22 & 3.21 & 403.18 & 30.97 & 0.809 & 0.761 & 0.915 \\
 48 & SC                         & 1 & 22 & 1.99 & 468.48 & 33.25 & 0.810 & 0.763 & 0.916 \\
 49 & SO                         & 3 & 24 & 1.55 & 518.11 & 31.14 & 0.789 & 0.752 & 0.941 \\
 50 & ClO                        & 2 & 25 & 1.33 & 579.64 & 30.57 & 0.803 & 0.762 & 0.931 \\
 51 & FCl                        & 1 & 26 & 0.91 & 607.86 & 29.71 & 0.813 & 0.766 & 0.915 \\
 52 & Si$_2$H$_6$                & 1 & 34 & 0.00 & 671.93 & 55.02 & 0.818 & 0.772 & 0.920 \\
 53 & CH$_3$Cl                   & 1 & 26 & 1.95 & 549.86 & 33.84 & 0.813 & 0.766 & 0.916 \\
 54 & H$_3$CSH                   & 1 & 26 & 1.55 & 494.08 & 36.34 & 0.812 & 0.765 & 0.916 \\
 55 & HOCl                       & 1 & 26 & 1.55 & 585.49 & 30.94 & 0.813 & 0.766 & 0.916 \\
 56 & SO$_2$                     & 1 & 32 & 1.71 & 655.49 & 41.21 & 0.817 & 0.771 & 0.918 \\
 57 & H                          & 2 &  1 & 0.00 &   0.50 &  2.41 & 0.760 & 0.714 & 0.925 \\
 58 & He                         & 1 &  2 & 0.00 &   2.87 &  1.87 & 0.779 & 0.725 & 0.906 \\
 59 & Li                         & 2 &  3 & 0.00 &   7.43 & 13.87 & 0.764 & 0.720 & 0.927 \\
 60 & Be                         & 1 &  4 & 0.00 &  14.59 & 13.05 & 0.783 & 0.731 & 0.909 \\
 61 & B                          & 2 &  5 & 0.00 &  24.57 & 12.61 & 0.766 & 0.722 & 0.927 \\
 62 & C                          & 3 &  6 & 0.00 &  37.76 & 11.06 & 0.749 & 0.710 & 0.939 \\
 63 & N                          & 4 &  7 & 0.00 &  54.49 &  9.68 & 0.731 & 0.698 & 0.946 \\
 64 & O                          & 3 &  8 & 0.00 &  75.09 &  9.06 & 0.752 & 0.713 & 0.938 \\
 65 & F                          & 2 &  9 & 0.00 &  99.53 &  8.28 & 0.772 & 0.727 & 0.926 \\
 66 & Ne                         & 1 & 10 & 0.00 & 128.66 &  7.55 & 0.791 & 0.738 & 0.907 \\
 67 & Na                         & 2 & 11 & 0.00 & 161.83 & 19.78 & 0.779 & 0.735 & 0.928 \\
 68 & Mg                         & 1 & 12 & 0.00 & 199.57 & 21.87 & 0.796 & 0.746 & 0.911 \\
 69 & Al                         & 2 & 13 & 0.00 & 241.93 & 26.04 & 0.784 & 0.741 & 0.929 \\
 70 & Si                         & 3 & 14 & 0.00 & 289.39 & 25.10 & 0.770 & 0.733 & 0.940 \\
 71 & P                          & 4 & 15 & 0.00 & 340.77 & 23.66 & 0.755 & 0.722 & 0.947 \\
 72 & S                          & 3 & 16 & 0.00 & 397.58 & 22.91 & 0.776 & 0.738 & 0.940 \\
 73 & Cl                         & 2 & 17 & 0.00 & 459.07 & 21.73 & 0.794 & 0.751 & 0.929 \\
 74 & Ar                         & 1 & 18 & 0.00 & 526.13 & 20.36 & 0.807 & 0.759 & 0.913 \\
 75 & PH                         & 3 & 16 & 0.40 & 346.54 & 24.90 & 0.774 & 0.737 & 0.940 \\
 76 & SH                         & 2 & 17 & 0.77 & 403.69 & 23.98 & 0.793 & 0.750 & 0.930 \\
 77 & H$^+$                      & 1 &  0 & 0.00 &   0.00 &  0.00 & 0.777 & 0.723 & 0.905 \\
 78 & He$^+$                     & 2 &  1 & 0.00 &   1.99 &  0.60 & 0.759 & 0.714 & 0.925 \\
 79 & Li$^+$                     & 1 &  2 & 0.00 &   7.22 &  0.70 & 0.779 & 0.725 & 0.905 \\
 80 & Be$^+$                     & 2 &  3 & 0.00 &  14.26 &  4.99 & 0.763 & 0.717 & 0.926 \\
 81 & B$^+$                      & 1 &  4 & 0.00 &  24.25 &  6.11 & 0.782 & 0.729 & 0.907 \\
 82 & C$^+$                      & 2 &  5 & 0.00 &  37.34 &  6.39 & 0.766 & 0.720 & 0.926 \\
 83 & N$^+$                      & 3 &  6 & 0.00 &  53.96 &  6.17 & 0.748 & 0.710 & 0.938 \\
 84 & O$^+$                      & 4 &  7 & 0.00 &  74.46 &  5.82 & 0.731 & 0.698 & 0.946 \\
 85 & F$^+$                      & 3 &  8 & 0.00 &  98.93 &  5.67 & 0.752 & 0.713 & 0.938 \\
 86 & Ne$^+$                     & 2 &  9 & 0.00 & 127.90 &  5.42 & 0.773 & 0.728 & 0.926 \\
 87 & Na$^+$                     & 1 & 10 & 0.00 & 161.63 &  5.09 & 0.791 & 0.739 & 0.907 \\
 88 & Mg$^+$                     & 2 & 11 & 0.00 & 199.29 & 10.60 & 0.778 & 0.734 & 0.927 \\
 89 & Al$^+$                     & 1 & 12 & 0.00 & 241.71 & 13.08 & 0.796 & 0.745 & 0.909 \\
 90 & Si$^+$                     & 2 & 13 & 0.00 & 288.61 & 15.64 & 0.784 & 0.740 & 0.928 \\
 91 & P$^+$                      & 3 & 14 & 0.00 & 340.42 & 16.27 & 0.770 & 0.733 & 0.940 \\
 92 & S$^+$                      & 4 & 15 & 0.00 & 397.23 & 16.07 & 0.756 & 0.723 & 0.947 \\
 93 & Cl$^+$                     & 3 & 16 & 0.00 & 458.63 & 16.14 & 0.777 & 0.739 & 0.940 \\
 94 & Ar$^+$                     & 2 & 17 & 0.00 & 525.55 & 15.13 & 0.795 & 0.752 & 0.929 \\
 95 & CH$_4$$^+$                 & 2 &  9 & 0.01 &  52.90 &  8.23 & 0.770 & 0.725 & 0.926 \\
 96 & NH$_3$$^+$                 & 2 &  9 & 0.00 &  67.71 &  7.32 & 0.771 & 0.725 & 0.926 \\
 97 & OH$^+$                     & 3 &  8 & 2.02 &  79.19 &  6.24 & 0.754 & 0.715 & 0.939 \\
 98 & OH$_2$$^+$                 & 2 &  9 & 2.12 &  84.52 &  6.77 & 0.774 & 0.729 & 0.926 \\
 99 & FH$^+$                     & 2 &  9 & 2.36 & 104.35 &  6.01 & 0.775 & 0.730 & 0.926 \\
100 & SiH$_4$$^+$                & 2 & 17 & 1.21 & 310.53 & 18.93 & 0.789 & 0.746 & 0.929 \\
101 & PH$^+$                     & 2 & 15 & 0.62 & 346.56 & 17.05 & 0.788 & 0.745 & 0.928 \\
102 & PH$_2$$^+$                 & 1 & 16 & 0.80 & 353.00 & 17.79 & 0.803 & 0.753 & 0.911 \\
103 & PH$_3$$^+$                 & 2 & 17 & 0.35 & 359.87 & 18.05 & 0.790 & 0.747 & 0.928 \\
104 & SH$^+$                     & 3 & 16 & 1.08 & 404.05 & 16.69 & 0.776 & 0.738 & 0.940 \\
105 & SH$_2$$^+$($^2$B$_1$)      & 2 & 17 & 1.37 & 411.17 & 17.29 & 0.793 & 0.750 & 0.929 \\
106 & SH$_2$$^+$($^2$A$_1$)      & 2 & 17 & 0.54 & 411.00 & 12.72 & 0.789 & 0.744 & 0.925 \\
107 & ClH$^+$                    & 2 & 17 & 1.53 & 466.10 & 16.58 & 0.794 & 0.751 & 0.929 \\
108 & C$_2$H$_2$$^+$             & 2 & 13 & 0.00 & 100.59 & 14.82 & 0.777 & 0.732 & 0.927 \\
109 & C$_2$H$_4$$^+$             & 2 & 15 & 0.00 & 110.31 & 15.16 & 0.779 & 0.734 & 0.926 \\
110 & CO$^+$                     & 2 & 13 & 2.73 & 135.32 & 12.48 & 0.781 & 0.736 & 0.927 \\
111 & N$_2$$^+$($^2$$\Sigma$$_g$)& 2 & 13 & 0.00 & 132.02 & 13.12 & 0.778 & 0.733 & 0.926 \\
112 & N$_2$$^+$($^2$$\Pi$$_u$)   & 2 & 13 & 0.00 & 130.77 & 13.95 & 0.777 & 0.731 & 0.924 \\
113 & O$_2$$^+$                  & 2 & 17 & 0.00 & 180.07 & 13.06 & 0.783 & 0.738 & 0.926 \\
114 & P$_2$$^+$                  & 2 & 29 & 0.00 & 741.28 & 33.44 & 0.808 & 0.767 & 0.932 \\
115 & S$_2$$^+$                  & 2 & 31 & 0.00 & 869.18 & 33.65 & 0.811 & 0.771 & 0.932 \\
116 & Cl$_2$$^+$                 & 2 & 33 & 0.00 & 997.85 & 33.40 & 0.814 & 0.773 & 0.933 \\
117 & FCl$^+$                    & 2 & 25 & 1.70 & 610.61 & 22.93 & 0.803 & 0.761 & 0.930 \\
118 & SC$^+$                     & 2 & 21 & 0.52 & 469.27 & 23.20 & 0.797 & 0.754 & 0.929 \\
119 & H$_3$$^+$                  & 1 &  2 & 0.00 &   3.06 &  1.50 & 0.779 & 0.725 & 0.905 \\
120 & C$_2$H$_3$$^+$             & 1 & 14 & 0.98 & 107.01 & 15.19 & 0.794 & 0.743 & 0.909 \\
121 & NH$_4$$^+$                 & 1 & 10 & 0.00 &  72.98 &  7.54 & 0.789 & 0.736 & 0.906 \\
122 & H$_3$O$^+$                 & 1 & 10 & 0.00 &  90.43 &  7.32 & 0.789 & 0.737 & 0.906 \\
123 & SiH$_5$$^+$                & 1 & 18 & 1.30 & 317.11 & 19.69 & 0.803 & 0.754 & 0.911 \\
124 & PH$_4$$^+$                 & 1 & 18 & 0.00 & 366.99 & 18.64 & 0.804 & 0.754 & 0.911 \\
125 & H$_3$S$^+$                 & 1 & 18 & 1.48 & 418.61 & 17.84 & 0.806 & 0.756 & 0.912 \\
126 & H$_2$Cl$^+$                & 1 & 18 & 1.90 & 473.92 & 16.99 & 0.807 & 0.758 & 0.912 \\
 \hline
\hline
 \label{table.1}
\end{longtable*}

$g_S$, $N_t$, D, ${\cal T}$ and Q of each molecule or atom in the
training set are listed in Table~\ref{table.1}. The conventional
B3LYP/6-311+G(3{\it df},2{\it p}) calculations are carried out to
evaluate AEs, IPs, PAs or TAEs of the molecules and atoms in the
training set, and the results are given in Tables
~\ref{table.2},~\ref{table.3},~\ref{table.4} and ~\ref{table.5},
respectively. Compared to the experimental data, the RMS
deviations are 3.0, 4.9, 1.6 and 10.3 kcal$\cdot$mol$^{-1}$ for
AEs, IPs, PAs and TAEs, respectively. The physical descriptors of
each molecule or atom in the training set are inputted to the
neural network, and the $\tilde{a}_0$, $\tilde{a}_X$ and
$\tilde{a}_C$ from the output layer are used to construct the
B3LYP functional which is used subsequently to calculate AE, IP,
PA or TAE. These values are then compared to the 116 energy values
in the training set, and the synaptic weights $\{ W_{ji}\}$ and
$\{ W'_{kj} \}$ are tuned accordingly. The final values of
synaptic weights are shown in Tables ~\ref{table.6} and
~\ref{table.7}. In Table ~\ref{table.8} we list the derivatives of
$\tilde{a}_0$, $\tilde{a}_X$ and $\tilde{a}_C$ with respect to
$x_i$ ($i$=0-5). The magnitude of a derivative reflects the
influence on $\tilde{a}_0$, $\tilde{a}_X$ and $\tilde{a}_C$ of the
corresponding physical descriptor. The larger the magnitude is,
the more significant the physical descriptor is to determine the
values of $\tilde{a}_0$, $\tilde{a}_X$ and $\tilde{a}_C$.
Derivatives in Table ~\ref{table.8} are obtained at $x_i=0.5$
($i$=1-5) and $x_0=1$. We find that the spin multiplicity $g_S$
and total kinetic energy ${\cal T}$ have the derivatives of the
largest two magnitudes. Similar results are observed at $x_i=0.1$
($i$=1-5) and $x_0=1$, or $x_i=0.9$ ($i$=1-5) and $x_0=1$. Hence
$g_s$ and ${\cal T}$ are identified as two most significant
descriptors to determine the high-order components of
$\tilde{a}_0$, $\tilde{a}_X$ and $\tilde{a}_C$. The final or
optimal values of $\tilde{a}_0$, $\tilde{a}_X$ and $\tilde{a}_C$
for each molecule or atom are listed in Table ~\ref{table.1}. Note
that their values are overall shifted from the original B3LYP
values, while the order
$\tilde{a}_X$$<$$\tilde{a}_0$$<$$\tilde{a}_C$ is kept for each
molecule or atom. This overall shift is caused by the finite basis
set. More importantly, their values are slightly different from
each other. Therefore, the resulting B3LYP functional is
system-dependent. We list the Neural-Networks-corrected AEs, IPs,
PAs and TAEs in Tables
~\ref{table.2},~\ref{table.3},~\ref{table.4} and ~\ref{table.5},
respectively. $\Delta_1$ and $\Delta_2$ in these tables are the
differences between the calculated values and the experimental
counterpart for the conventional B3LYP/6-311+G(3{\it df},2{\it p})
and the Neural-Networks-based B3LYP/6-311+G(3{\it df},2{\it p})
calculations, respectively. Compared to their experimental
counterparts, the RMS deviations of Neural-Networks-based
B3LYP/6-311+G(3{\it df},2{\it p}) calculations are 2.4, 3.7, 1.6
and 2.7 kcal$\cdot$mol$^{-1}$ for AE, IP, PA and TAE,
respectively, and are less than those of the conventional
B3LYP/6-311+G(3{\it df},2{\it p}) calculations~(cf. Table
~\ref{table.9}). We note that the Neural-Networks-based
B3LYP/6-311+G(3{\it df},2{\it p}) calculations yield much improved
TAE results (see Table ~\ref{table.5}). In Becke's original
work~\cite{b3lyp}, the RMS deviations are 2.9, 3.9, 1.9, and 4.1
kcal$\cdot$mol$^{-1}$ for AE, IP, PA and TAE, respectively. The
new B3LYP/6-311+G(3{\it df},2{\it p}) calculations yield improved
results in comparison to Becke's work~\cite{b3lyp} (cf. Table
~\ref{table.9}).

\begin{table}
\caption{Atomization Energy (kcal$\cdot$mol$^{-1}$)}
 \begin{ruledtabular}
 \begin{tabular}{rlrrrrr}
  No. & Name &Expt. & DFT-1\footnotemark[1]
    & $~~~\Delta_1$  & DFT-NN\footnotemark[2] & $~~~\Delta_2$\\
\hline
 1 & H$_2$             & 103.5  & 103.9 &  0.4 & 103.8 &  0.3 \\
 2 & LiH               &  56.0  &  56.4 &  0.4 &  56.4 &  0.4 \\
 3 & BeH               &  46.9  &  55.0 &  8.1 &  53.8 &  6.9 \\
 4 & CH                &  79.9  &  81.4 &  1.5 &  80.8 &  0.9 \\
 5 & CH$_2$($^3$B$_1$) & 179.6  & 181.4 &  1.8 & 179.8 &  0.2 \\
 6 & CH$_2$($^1$A$_1$) & 170.6  & 170.4 & -0.2 & 170.7 &  0.1 \\
 7 & CH$_3$            & 289.2  & 291.3 &  2.1 & 289.7 &  0.5 \\
 8 & CH$_4$            & 392.5  & 392.9 &  0.4 & 392.9 &  0.4 \\
 9 & NH                &  79.0  &  83.4 &  4.4 &  81.5 &  2.5 \\
10 & NH$_2$            & 170.0  & 176.0 &  6.0 & 173.8 &  3.8 \\
11 & NH$_3$            & 276.7  & 279.5 &  2.8 & 278.4 &  1.7 \\
12 & OH                & 101.3  & 102.9 &  1.6 & 101.8 &  0.5 \\
13 & OH$_2$            & 219.3  & 217.6 & -1.7 & 218.4 & -0.9 \\
14 & FH                & 135.2  & 133.5 & -1.7 & 134.5 & -0.7 \\
15 & Li$_2$            &  24.0  &  20.5 & -3.5 &  21.2 & -2.8 \\
16 & LiF               & 137.6  & 135.8 & -1.8 & 137.0 & -0.6 \\
17 & C$_2$H$_2$        & 388.9  & 386.4 & -2.5 & 387.1 & -1.8 \\
18 & C$_2$H$_4$        & 531.9  & 531.3 & -0.6 & 532.5 &  0.6 \\
19 & C$_2$H$_6$        & 666.3  & 664.8 & -1.5 & 665.0 & -1.3 \\
20 & CN                & 176.6  & 176.7 &  0.1 & 174.5 & -2.1 \\
21 & HCN               & 301.8  & 303.6 &  1.8 & 304.1 &  2.3 \\
22 & CO                & 256.2  & 253.1 & -3.1 & 254.7 & -1.5 \\
23 & HCO               & 270.3  & 273.2 &  2.9 & 272.0 &  1.7 \\
24 & H$_2$CO           & 357.2  & 357.8 &  0.6 & 358.6 &  1.4 \\
25 & H$_3$COH          & 480.8  & 480.0 & -0.8 & 481.3 &  0.5 \\
26 & N$_2$             & 225.1  & 226.1 &  1.0 & 226.0 &  0.9 \\
27 & H$_2$NNH$_2$      & 405.4  & 410.9 &  5.5 & 410.0 &  4.6 \\
28 & NO                & 150.1  & 153.0 &  2.9 & 150.7 &  0.6 \\
29 & O$_2$             & 118.0  & 121.7 &  3.7 & 117.7 & -0.3 \\
30 & HOOH              & 252.3  & 249.9 & -2.4 & 252.4 &  0.1 \\
31 & F$_2$             &  36.9  &  34.8 & -2.1 &  38.6 &  1.7 \\
32 & CO$_2$            & 381.9  & 382.4 &  0.5 & 385.5 &  3.6 \\
33 & SiH$_2$($^1$A$_1$)& 144.4  & 146.2 &  1.8 & 146.4 &  2.0 \\
34 & SiH$_2$($^3$B$_1$)& 123.4  & 125.4 &  2.0 & 122.8 & -0.6 \\
35 & SiH$_3$           & 214.0  & 210.6 & -3.4 & 207.4 & -6.6 \\
36 & SiH$_4$           & 302.8  & 303.9 &  1.1 & 303.4 &  0.6 \\
37 & PH$_2$            & 144.7  & 150.5 &  5.8 & 147.9 &  3.2 \\
38 & PH$_3$            & 227.4  & 230.2 &  2.8 & 228.8 &  1.4 \\
39 & SH$_2$            & 173.2  & 172.6 & -0.6 & 172.8 & -0.4 \\
40 & ClH               & 102.2  & 101.1 & -1.1 & 101.7 & -0.5 \\
41 & Na$_2$            &  16.6  &  17.1 &  0.5 &  21.0 &  4.4 \\
42 & Si$_2$            &  74.0  &  70.1 & -3.9 &  69.6 & -4.4 \\
43 & P$_2$             & 116.1  & 115.5 & -0.6 & 115.1 & -1.0 \\
44 & S$_2$             & 100.7  & 102.0 &  1.3 & 103.6 &  2.9 \\
45 & Cl$_2$            &  57.2  &  54.4 & -2.8 &  56.9 & -0.3 \\
46 & NaCl              &  97.5  &  93.1 & -4.4 &  99.0 &  1.5 \\
47 & SiO               & 190.5  & 185.6 & -4.9 & 188.0 & -2.5 \\
48 & SC                & 169.5  & 164.7 & -4.8 & 168.1 & -1.4 \\
49 & SO                & 123.5  & 124.8 &  1.3 & 124.4 &  0.9 \\
50 & ClO               &  63.3  &  65.1 &  1.8 &  67.1 &  3.8 \\
51 & FCl               &  60.3  &  59.5 & -0.8 &  65.6 &  5.3 \\
52 & Si$_2$H$_6$       & 500.1  & 499.3 & -0.8 & 496.9 & -3.2 \\
53 & CH$_3$Cl          & 371.0  & 369.6 & -1.4 & 371.8 &  0.8 \\
54 & H$_3$CSH          & 445.1  & 443.0 & -2.1 & 444.7 & -0.4 \\
55 & HOCl              & 156.3  & 154.8 & -1.5 & 159.1 &  2.8 \\
56 & SO$_2$            & 254.0  & 246.4 & -7.6 & 253.3 & -0.7 \\
   \end{tabular}
 \label{table.2}
     \end{ruledtabular}
\footnotetext[1]{conventional B3LYP/6-311+G(3{\it df},2{\it p})}
\footnotetext[2]{Neural-Networks-based B3LYP/6-311+G(3{\it
df},2{\it p})}
\end{table}

\begin{table}
 \caption{Ionization Potential (eV)}
  \begin{ruledtabular}
  \begin{center}
   \begin{tabular}{rlrrrrr}
No. & Name & Expt. & DFT-1\footnote{conventional
B3LYP/6-311+G(3{\it df},2{\it p})}
    & $~~~\Delta_1$  & DFT-NN\footnote{Neural-Networks-based B3LYP/6-311+G(3{\it df},2{\it p})} & $~~~\Delta_2$
\\\hline
 1 & H                       & 13.60 & 13.66 &  0.06 & 13.58 & -0.02 \\
 2 & He                      & 24.59 & 24.93 &  0.34 & 24.82 &  0.23 \\
 3 & Li                      &  5.39 &  5.62 &  0.23 &  5.53 &  0.14 \\
 4 & Be                      &  9.32 &  9.12 & -0.20 &  9.06 & -0.26 \\
 5 & B                       &  8.30 &  8.74 &  0.44 &  8.64 &  0.34 \\
 6 & C                       & 11.26 & 11.55 &  0.29 & 11.44 &  0.18 \\
 7 & N                       & 14.54 & 14.67 &  0.13 & 14.56 &  0.02 \\
 8 & O                       & 13.61 & 14.16 &  0.55 & 13.95 &  0.34 \\
 9 & F                       & 17.42 & 17.76 &  0.34 & 17.62 &  0.20 \\
10 & Ne                      & 21.56 & 21.77 &  0.21 & 21.69 &  0.13 \\
11 & Na                      &  5.14 &  5.42 &  0.28 &  5.27 &  0.13 \\
12 & Mg                      &  7.65 &  7.73 &  0.08 &  7.72 &  0.07 \\
13 & Al                      &  5.98 &  6.02 &  0.04 &  5.88 & -0.10 \\
14 & Si                      &  8.15 &  8.11 & -0.04 &  8.08 & -0.07 \\
15 & P                       & 10.49 & 10.38 & -0.11 & 10.31 & -0.18 \\
16 & S                       & 10.36 & 10.55 &  0.19 & 10.32 & -0.04 \\
17 & Cl                      & 12.97 & 13.07 &  0.10 & 12.95 & -0.02 \\
18 & Ar                      & 15.76 & 15.80 &  0.04 & 15.82 &  0.06 \\
19 & CH$_4$                  & 12.62 & 12.46 & -0.16 & 12.47 & -0.15 \\
20 & NH$_3$                  & 10.18 & 10.20 &  0.02 & 10.20 &  0.02 \\
21 & OH                      & 13.01 & 13.23 &  0.22 & 13.12 &  0.11 \\
22 & OH$_2$                  & 12.62 & 12.62 &  0.00 & 12.61 & -0.01 \\
23 & FH                      & 16.04 & 16.10 &  0.06 & 16.07 &  0.03 \\
24 & SiH$_4$                 & 11.00 & 10.91 & -0.09 & 10.93 & -0.07 \\
25 & PH                      & 10.15 & 10.17 &  0.02 & 10.08 & -0.07 \\
26 & PH$_2$                  &  9.82 &  9.92 &  0.10 &  9.77 & -0.05 \\
27 & PH$_3$                  &  9.87 &  9.83 & -0.04 &  9.81 & -0.06 \\
28 & SH                      & 10.37 & 10.46 &  0.09 & 10.33 & -0.04 \\
29 & SH$_2$($^2$B$_1$)       & 10.47 & 10.41 & -0.06 & 10.38 & -0.09 \\
30 & SH$_2$($^2$A$_1$)       & 12.78 & 12.65 & -0.13 & 12.63 & -0.15 \\
31 & ClH                     & 12.75 & 12.74 & -0.01 & 12.69 & -0.06 \\
32 & C$_2$H$_2$              & 11.40 & 11.25 & -0.15 & 11.30 & -0.10 \\
33 & C$_2$H$_4$              & 10.51 & 10.29 & -0.22 & 10.37 & -0.14 \\
34 & CO                      & 14.01 & 14.18 &  0.17 & 14.24 &  0.23 \\
35 & N$_2$($^2$$\Sigma$$_g$) & 15.58 & 15.84 &  0.26 & 15.93 &  0.35 \\
36 & N$_2$($^2$$\Pi$$_u$)    & 16.70 & 16.66 & -0.04 & 16.71 &  0.01 \\
37 & O$_2$                   & 12.07 & 12.58 &  0.51 & 12.44 &  0.37 \\
38 & P$_2$                   & 10.53 & 10.34 & -0.19 & 10.35 & -0.18 \\
39 & S$_2$                   &  9.36 &  9.55 &  0.19 &  9.37 &  0.01 \\
40 & Cl$_2$                  & 11.50 & 11.38 & -0.12 & 11.38 & -0.12 \\
41 & FCl                     & 12.66 & 12.62 & -0.04 & 12.72 &  0.06 \\
42 & SC                      & 11.33 & 11.43 &  0.10 & 11.51 &  0.18 \\
   \end{tabular}
  \end{center}
   \end{ruledtabular}
 \label{table.3}
\end{table}

\begin{table}
 \caption{Proton Affinity (kcal$\cdot$mol$^{-1}$)}
  \begin{ruledtabular}
  \begin{center}
   \begin{tabular}{rlrrrrr}
No. & Name & Expt. & DFT-1\footnote{conventional
B3LYP/6-311+G(3{\it df},2{\it p})}
    & $~~~\Delta_1$  & DFT-NN\footnote{Neural-Networks-based B3LYP/6-311+G(3{\it df},2{\it p})} & $~~~\Delta_2$
\\\hline
1 & H$_2$      & 100.8 &  98.6 & -2.2 &  98.4 & -2.4 \\
2 & C$_2$H$_2$ & 152.3 & 154.0 &  1.7 & 154.3 &  2.0 \\
3 & NH$_3$     & 202.5 & 201.4 & -1.1 & 201.6 & -0.9 \\
4 & H$_2$O     & 165.1 & 162.1 & -3.0 & 162.3 & -2.8 \\
5 & SiH$_4$    & 154.0 & 153.2 & -0.8 & 154.3 &  0.3 \\
6 & PH$_3$     & 187.1 & 186.0 & -1.1 & 185.7 & -1.4 \\
7 & H$_2$S     & 168.8 & 168.2 & -0.6 & 167.9 & -0.9 \\
8 & HCl        & 133.6 & 132.8 & -0.8 & 133.9 &  0.3 \\
   \end{tabular}
  \end{center}
   \end{ruledtabular}
 \label{table.4}
\end{table}

\begin{table}
 \caption{Total Atomic Energy (hartrees)}
  \begin{ruledtabular}
  \begin{center}
   \begin{tabular}{rlrrrrr}
No. & Name & Expt. & DFT-1\footnote{conventional
B3LYP/6-311+G(3{\it df},2{\it p})}
    & $\Delta_1$  & DFT-NN\footnote{Neural-Networks-based B3LYP/6-311+G(3{\it df},2{\it p})} & $\Delta_2$
\\\hline
 1 & H  &   -0.500 &   -0.502 & -0.002 &   -0.499 &  0.001 \\
 2 & He &   -2.904 &   -2.913 & -0.009 &   -2.906 & -0.002 \\
 3 & Li &   -7.478 &   -7.491 & -0.013 &   -7.482 & -0.004 \\
 4 & Be &  -14.667 &  -14.671 & -0.004 &  -14.661 &  0.006 \\
 5 & B  &  -24.654 &  -24.663 & -0.009 &  -24.649 &  0.005 \\
 6 & C  &  -37.845 &  -37.857 & -0.012 &  -37.841 &  0.004 \\
 7 & N  &  -54.590 &  -54.601 & -0.011 &  -54.583 &  0.007 \\
 8 & O  &  -75.067 &  -75.091 & -0.024 &  -75.069 & -0.002 \\
 9 & F  &  -99.731 &  -99.762 & -0.031 &  -99.737 & -0.006 \\
10 & Ne & -128.937 & -128.960 & -0.023 & -128.935 &  0.002 \\
   \end{tabular}
  \end{center}
   \end{ruledtabular}
 \label{table.5}
\end{table}

\begin{table}
 \begin{ruledtabular}
 \caption{ Optimized Synaptic Weights W$_{ji}$ }
  \begin{tabular}{rrrrrrr}
      &&& {\it j}=1 &&& {\it j}=2  \\\hline
  W$_{j1}$ &&& -0.89 &&&  1.11  \\
  W$_{j2}$ &&&  0.52 &&& -0.09  \\
  W$_{j3}$ &&&  0.18 &&&  0.09  \\
  W$_{j4}$ &&&  0.78 &&&  0.20  \\
  W$_{j5}$ &&&  0.22 &&&  0.28  \\
  W$_{j0}$ &&&  0.15 &&&  0.06  \\
 \end{tabular}
  \end{ruledtabular}
  \label{table.6}
   \end{table}

\begin{table}
 \begin{ruledtabular}
 \caption{ Optimized Synaptic Weights W'$_{kj}$ }
  \begin{tabular}{rrrrrrrrrr}
      &&& {\it k}=1 &&& {\it k}=2 &&& {\it k}=3 \\\hline
  W'$_{k1}$ &&& 0.21 &&&-0.02 &&& 0.46 \\
  W'$_{k2}$ &&& 0.18 &&& 0.06 &&& 0.36 \\
  W'$_{k0}$ &&&-0.03 &&& 0.54 &&& 0.53 \\
 \end{tabular}
  \end{ruledtabular}
  \label{table.7}
   \end{table}

\begin{table}
 \begin{ruledtabular}
 \caption{ The Derivatives of $\tilde{a}_0$, $\tilde{a}_X$ and $\tilde{a}_C$
{\it w.r.t.} Each Physical Descriptor$^a$}
  \begin{tabular}{rrrrrrrrrr}
      &&& \large{$\partial\tilde{a}_0\over\partial{x_i}$} &&& \large{$\partial\tilde{a}_X\over\partial{x_i}$}
      &&& \large{$\partial\tilde{a}_C\over\partial{x_i}$} \\\hline
  {\it i}=1 &&& -0.067 &&& -0.036 &&&  0.099 \\
  {\it i}=2 &&&  0.035 &&&  0.034 &&& -0.010 \\
  {\it i}=3 &&&  0.011 &&&  0.015 &&&  0.007 \\
  {\it i}=4 &&&  0.050 &&&  0.058 &&&  0.014 \\
  {\it i}=5 &&&  0.012 &&&  0.022 &&&  0.023 \\
  {\it i}=0 &&&  0.009 &&&  0.011 &&&  0.044 \\
 \end{tabular}
  \label{table.8}
   \end{ruledtabular}
   \footnotetext[1]{Derivatives are obtained at $x_i$=0.5 ({\it i}=1-5)
and $x_0$=1.}
   \end{table}

\begin{table}
 \caption{RMS (all data are in the unit of kcal$\cdot$mol$^{-1}$)}
 \begin{ruledtabular}
   \begin{tabular}{rllrrrr}
 Properties && AE  & IP & PA & TAE & Overall \\
 Number of samples && 56 & 42 & 8 & 10 & 116\\
 \hline
A\footnote{Becke's work} && 2.9 & 3.9 & 1.9  & 4.1 & 3.4 \\
DFT-1\footnote{conventional B3LYP/6-311+G(3{\it df},2{\it p})} && 3.0 & 4.9 & 1.6 & 10.3 & 4.7 \\
DFT-NN\footnote{Neural-Networks-based B3LYP/6-311+G(3{\it
df},2{\it p})}
 && 2.4 & 3.7 & 1.6 & 2.7 & 2.9 \\
   \end{tabular}
 \label{table.9}
\end{ruledtabular}
\end{table}
To examine the performance of our neural network, a test is
carried out by calculating the IPs of 24 molecules which are
selected from the G2 test set~\cite{g3}. To save the computational
time, only the 24 smallest molecules are selected besides those
appeared in the training set and are termed as the testing set.
Physical descriptors of each molecule in the testing set are
inputted into our neural network and the Neural-Networks-corrected
$\tilde{a}_0$, $\tilde{a}_X$ and $\tilde{a}_C$ are used to
construct the new B3LYP functional (see Table ~\ref{table.10}). To
calculate their IPs, the cation counterparts of the 24 molecules
need to be included as well. Their $\tilde{a}_0$, $\tilde{a}_X$
and $\tilde{a}_C$ are also listed in Table ~\ref{table.10}. The
resulting IPs are given in Table ~\ref{table.11}, in comparison to
those obtained from the conventional B3LYP/6-311+G(3{\it df},2{\it
p}) calculations. Obviously, the resulting IPs for most molecules
are improved upon the Neural-Networks correction. For the
Neural-Networks-based B3LYP/6-311+G(3{\it df},2{\it p})
calculation, its RMS deviation for the 24 molecules is reduced to
2.2 kcal$\cdot$mol$^{-1}$ from the original 3.6
kcal$\cdot$mol$^{-1}$. This test demonstrates the validity of our
Neural-Networks-based functional.

\begin{table*}
 \caption{Descriptors and Parameters of Testing Set}
    \begin{ruledtabular}
   \begin{tabular}{rlcrrrrccc}
No. & Name & g$_S$ & Nt &~D~ & $\cal T$~~ &  Q~~~ & $\tilde{a}_0$ & $\tilde{a}_X$ & $\tilde{a}_C$\\
            &      &     &    &~(DB)   & ~~(a.u.) &  ~~(DB$\cdot$\AA)  &       &       &       \\
     \hline
 1 & CF$_2$          & 1 & 24 & 0.51 & 301.68 & 27.87 & 0.805 & 0.754 & 0.909 \\
 2 & CH$_2$          & 3 &  8 & 0.63 &  45.13 & 12.98 & 0.754 & 0.714 & 0.936 \\
 3 & CH$_2$S         & 1 & 24 & 1.75 & 481.58 & 33.56 & 0.810 & 0.761 & 0.913 \\
 4 & CH$_3$Cl        & 1 & 26 & 1.96 & 549.86 & 33.85 & 0.812 & 0.763 & 0.914 \\
 5 & CH$_3$F         & 1 & 18 & 1.87 & 176.46 & 21.21 & 0.798 & 0.746 & 0.908 \\
 6 & CH$_3$          & 2 &  9 & 0.00 &  49.28 & 13.87 & 0.771 & 0.725 & 0.924 \\
 7 & CH$_3$OH        & 1 & 18 & 1.67 & 155.45 & 22.90 & 0.798 & 0.746 & 0.908 \\
 8 & CH$_3$O         & 2 & 17 & 2.11 & 149.00 & 22.23 & 0.784 & 0.739 & 0.926 \\
 9 & CHO             & 2 & 15 & 1.69 & 140.01 & 19.87 & 0.781 & 0.736 & 0.925 \\
10 & CO$_2$          & 1 & 22 & 0.00 & 246.26 & 28.71 & 0.803 & 0.752 & 0.909 \\
11 & COS             & 1 & 30 & 0.85 & 589.92 & 41.32 & 0.815 & 0.767 & 0.915 \\
12 & HOF             & 1 & 18 & 1.95 & 208.68 & 17.89 & 0.799 & 0.746 & 0.907 \\
13 & NH$_2$          & 2 &  9 & 1.83 &  63.22 & 12.11 & 0.772 & 0.726 & 0.924 \\
14 & NH              & 3 &  8 & 1.54 &  58.63 & 10.94 & 0.754 & 0.715 & 0.936 \\
15 & SC              & 1 & 22 & 1.92 & 468.67 & 33.28 & 0.809 & 0.760 & 0.914 \\
16 & B$_2$H$_4$      & 1 & 14 & 0.79 &  76.44 & 26.08 & 0.793 & 0.742 & 0.909 \\
17 & C$_2$H$_5$      & 2 & 17 & 0.34 & 115.61 & 25.37 & 0.782 & 0.738 & 0.926 \\
18 & CH$_3$SH        & 1 & 26 & 1.54 & 494.06 & 36.34 & 0.811 & 0.763 & 0.914 \\
19 & CS$_2$          & 1 & 38 & 0.00 & 942.06 & 53.04 & 0.822 & 0.776 & 0.920 \\
20 & N$_2$H$_2$      & 1 & 16 & 0.00 & 142.57 & 21.13 & 0.795 & 0.743 & 0.907 \\
21 & N$_2$H$_3$      & 2 & 17 & 2.56 & 147.28 & 21.96 & 0.784 & 0.739 & 0.926 \\
22 & Si$_2$H$_2$     & 1 & 30 & 0.57 & 643.83 & 50.42 & 0.816 & 0.770 & 0.918 \\
23 & Si$_2$H$_4$     & 1 & 32 & 0.00 & 658.51 & 51.48 & 0.817 & 0.771 & 0.918 \\
24 & SiH$_3$         & 2 & 17 & 0.07 & 306.32 & 27.95 & 0.789 & 0.745 & 0.928 \\
25 & CF$_2$$^+$      & 2 & 23 & 1.08 & 303.37 & 20.16 & 0.792 & 0.747 & 0.925 \\
26 & CH$_2$$^+$      & 2 &  7 & 0.52 &  44.66 &  7.41 & 0.768 & 0.721 & 0.923 \\
27 & CH$_2$S$^+$     & 2 & 23 & 1.70 & 481.65 & 24.05 & 0.798 & 0.754 & 0.927 \\
28 & CH$_3$Cl$^+$    & 2 & 25 & 1.89 & 550.53 & 24.57 & 0.801 & 0.757 & 0.927 \\
29 & CH$_3$F$^+$     & 2 & 17 & 3.72 & 177.34 & 13.46 & 0.784 & 0.738 & 0.924 \\
30 & CH$_3$$^+$      & 1 &  8 & 0.00 &  48.83 &  7.61 & 0.784 & 0.730 & 0.903 \\
31 & CH$_3$OH$^+$    & 2 & 17 & 1.43 & 155.82 & 14.56 & 0.782 & 0.736 & 0.924 \\
32 & CH$_3$O$^+$     & 3 & 16 & 2.44 & 149.57 & 14.24 & 0.766 & 0.726 & 0.936 \\
33 & CHO$^+$         & 1 & 14 & 3.76 & 140.98 & 12.86 & 0.794 & 0.741 & 0.906 \\
34 & CO$_2$$^+$      & 2 & 21 & 0.00 & 245.22 & 20.89 & 0.788 & 0.743 & 0.925 \\
35 & COS$^+$         & 2 & 29 & 1.66 & 588.35 & 31.06 & 0.804 & 0.761 & 0.928 \\
36 & HOF$^+$         & 2 & 17 & 2.80 & 210.63 & 12.46 & 0.784 & 0.738 & 0.924 \\
37 & NH$_2$$^+$      & 3 &  8 & 0.56 &  62.72 &  7.39 & 0.753 & 0.713 & 0.935 \\
38 & NH$^+$          & 2 &  7 & 1.73 &  58.03 &  6.69 & 0.769 & 0.722 & 0.923 \\
39 & SC$^+$          & 2 & 21 & 0.54 & 469.27 & 23.22 & 0.795 & 0.751 & 0.927 \\
40 & B$_2$H$_4$$^+$  & 2 & 13 & 0.28 &  75.62 & 16.76 & 0.776 & 0.730 & 0.924 \\
41 & C$_2$H$_5$$^+$  & 1 & 16 & 0.70 & 117.00 & 16.11 & 0.794 & 0.741 & 0.905 \\
42 & CH$_3$SH$^+$    & 2 & 25 & 1.16 & 494.52 & 25.68 & 0.799 & 0.755 & 0.927 \\
43 & CS$_2$$^+$      & 2 & 37 & 0.00 & 941.86 & 40.10 & 0.815 & 0.773 & 0.930 \\
44 & N$_2$H$_2$$^+$  & 2 & 15 & 0.00 & 143.30 & 13.30 & 0.779 & 0.733 & 0.923 \\
45 & N$_2$H$_3$$^+$  & 1 & 16 & 2.55 & 148.94 & 14.25 & 0.795 & 0.743 & 0.905 \\
46 & Si$_2$H$_2$$^+$ & 2 & 29 & 0.22 & 642.41 & 34.76 & 0.806 & 0.763 & 0.929 \\
47 & Si$_2$H$_4$$^+$ & 2 & 31 & 0.00 & 656.32 & 36.04 & 0.807 & 0.764 & 0.929 \\
48 & SiH$_3$$^+$     & 2 & 16 & 0.06 & 306.15 & 17.96 & 0.786 & 0.741 & 0.925 \\
   \end{tabular}
   \end{ruledtabular}
 \label{table.10}
\end{table*}

\begin{table}
 \caption{Ionization Potential of Testing Set (all data are in unit of eV)}
    \begin{ruledtabular}
   \begin{tabular}{rlrrrrr}
No. & Name & Expt. & DFT-1\footnote{RMS = 3.6
kcal$\cdot$mol$^{-1}$}  & ~~~~$\Delta_1$
    & DFT-NN\footnote{RMS = 2.2 kcal$\cdot$mol$^{-1}$}  & ~~~~$\Delta_2$\\\hline
 1 & CF$_2$       & 11.42 & 11.35 & -0.07 & 11.48 &  0.06  \\
 2 & CH$_2$       & 10.39 & 10.40 &  0.01 & 10.28 & -0.11  \\
 3 & CH$_2$S      &  9.38 &  9.28 & -0.10 &  9.34 & -0.04  \\
 4 & CH$_3$Cl     & 11.22 & 11.08 & -0.14 & 11.10 & -0.12  \\
 5 & CH$_3$F      & 12.47 & 12.30 & -0.17 & 12.38 & -0.09  \\
 6 & CH$_3$       &  9.84 &  9.95 &  0.09 &  9.76 & -0.08  \\
 7 & CH$_3$OH     & 10.85 & 10.59 & -0.26 & 10.68 & -0.17  \\
 8 & CH$_3$O      & 10.72 & 10.58 & -0.14 & 10.55 & -0.17  \\
 9 & CHO          &  8.14 &  8.49 &  0.35 &  8.27 &  0.13  \\
10 & CO$_2$       & 13.77 & 13.74 & -0.03 & 13.86 &  0.09  \\
11 & COS          & 11.18 & 11.19 &  0.01 & 11.25 &  0.07  \\
12 & HOF          & 12.71 & 12.66 & -0.05 & 12.73 &  0.02  \\
13 & NH$_2$       & 11.14 & 11.33 &  0.19 & 11.22 &  0.08  \\
14 & NH           & 13.49 & 13.68 &  0.19 & 13.58 &  0.09  \\
15 & SC           & 11.33 & 11.43 &  0.10 & 11.44 &  0.11  \\
16 & B$_2$H$_4$   &  9.70 &  9.50 & -0.20 &  9.54 & -0.16  \\
17 & C$_2$H$_5$   &  8.12 &  8.22 &  0.10 &  7.97 & -0.15  \\
18 & CH$_3$SH     &  9.44 &  9.33 & -0.11 &  9.43 & -0.01  \\
19 & CS$_2$       & 10.07 & 10.03 & -0.04 & 10.03 & -0.04  \\
20 & N$_2$H$_2$   &  9.59 &  9.55 & -0.04 &  9.62 &  0.03  \\
21 & N$_2$H$_3$   &  7.61 &  7.90 &  0.29 &  7.57 & -0.04  \\
22 & Si$_2$H$_2$  &  8.20 &  8.03 & -0.17 &  8.15 & -0.05  \\
23 & Si$_2$H$_4$  &  8.09 &  7.90 & -0.21 &  8.04 & -0.05  \\
24 & SiH$_3$      &  8.14 &  8.18 &  0.04 &  8.06 & -0.08  \\
   \end{tabular}
   \end{ruledtabular}
 \label{table.11}
\end{table}
\section{Discussion and Conclusion}

There are currently two schools of density functional
construction: the reductionist school and the semiempiricist
school. The reductionists attempt to deduce the universal
exchange-correlation functional from the first-principles. The
Jacob's ladder~\cite{perdew00} of density functional
approximations depicts the approach that the reductionists take
towards the universal exchange-correlation functional of chemical
accuracy. Becke realized that the existence and uniqueness of
exact exchange-correlation functional do not guarantee that the
functional is expressible in simple or even not so-simple
analytical form, and introduced the semiempirical approach to
construct accurate exchange-correlation functionals. We go beyond
the semiempirical approach by constructing the
Neural-Networks-based exchange-correlation functional. Our
generalized functional is a neural network whose structure and
synaptic weights are determined by accurate experimental data. It
is dynamic, and evolves readily when more accurate experimental
data become available. Although the parameters in the resulting
functional, such as $\tilde{a}_0$, $\tilde{a}_X$ and
$\tilde{a}_C$, are system-dependent as compared to the universal
functionals adopted by both reductionists and semiempiricists, the
neural network is not system-dependent and is regarded as a
generalized universal functional. Our approach relies on Neural
Networks to discover automatically the hidden regularities or
rules from large amount of experimental data. It is thus distinct
from the semiempirical approach. We term it as the discovery
approach. Compared to the conventional B3LYP/6-311+G(3{\it
df},2{\it p}) calculations, the Neural-Networks-based
B3LYP/6-311+G(3{\it df},2{\it p}) calculations yield much improved
AEs, IPs, PAs and TAEs (cf. Table ~\ref{table.9}). However, the
improvement over Becke's calculation~\cite{b3lyp} is not as
significant. This leaves room for further improvement or
investigation.

To summarize, we have developed a promising new approach, the
Neural-Networks-based approach, to construct the accurate DFT
exchange-correlation functional. The improved B3LYP functional
developed in this work is certainly not yet the final
exchange-correlation functional of chemical accuracy that we seek
for. Our work opens the door of an entirely different methodology
to develop the accurate exchange-correlation functionals. The
Neural-Networks-based functional can be systematically improved as
more or better experimental data become available. The
introduction of Neural Networks to the construction of
exchange-correlation functionals is potentially a powerful tool in
computational chemistry and physics, and may open the possibility
for first-principles methods being employed routinely as
predictive tools in materials research and development.

We thank Prof. YiJing Yan for extensive discussion on the subject.
Support from the Hong Kong Research Grant Council (RGC) and the
Committee for Research and Conference Grants (CRCG) of the
University of Hong Kong is gratefully acknowledged.

\end{document}